\documentclass[a4paper,11pt]{article}
\pdfoutput=1 % if your are submitting a pdflatex (i.e. if you have
\usepackage{jheppub} % for details on the use of the package, please
\usepackage[T1]{fontenc} % if needed
\usepackage{braket}
\usepackage{xcolor}

\title{\boldmath Measuring operator size growth in quantum quench experiments}

\author[a,b]{Xiao-Liang Qi,}
\author[b]{Emily J. Davis,}
\author[b]{Avikar Periwal,}
\author[b]{and Monika Schleier-Smith}
\affiliation[a]{Stanford Institute for Theoretical Physics, Stanford University}
\affiliation[b]{Department of Physics, Stanford University}
\abstract{Operator scrambling denotes the evolution of a simple operator into a complicated one (in the Heisenberg picture), which characterizes quantum chaos in many-body systems. More specifically, a simple operator evolves into a linear superposition of many operators, most of which are many-body operators supported on a region of size much larger than $1$. In general, an operator does not have a definite size but is characterized by a probability distribution of size. The operator size is related to out-of-time-order correlation functions, but these are generically difficult to obtain from experimental observables. In this paper we show that the operator size distribution can be measured in quantum quench experiments. In a quantum spin system, we propose to prepare an ensemble of initial states which are direct product states of random pure states of each spin qudit, and measure a simple physical observable (such as a particular component of spin) at later time $t$. The initial state dependence of the expectation value measures a particular component of the operator size distribution. Furthermore, many other features of the operator size distribution can be measured by analyzing the same data, such as the support of the operator in space.}

\begin{document} 
\maketitle
\flushbottom

\section{Introduction}

In recent years, progress has been made in understanding and characterizing quantum chaos in many-body systems. In the research of black hole thermodynamics and holographic duality\cite{maldacena1999large}, it was realized that black holes are highly chaotic systems\cite{hayden2007black,sekino2008fast}. The intuition from black hole physics helped to introduce new characteristics of many-body chaos, such as the out-of-time-ordered correlation function (OTOC)\cite{shenker2014black,shenker2014multiple,kitaev2014talk} (although the latter has been studied historically in disordered superconductors\cite{larkin1969quasiclassical}). Compared with time-ordered correlation functions that contribute to physical response functions, the OTOC probes the ``operator scrambling", which refers to the fact that a simple local operator evolves in Heisenberg evolution into a complicated multi-body operator, instead of just a linear superposition of local operators at different location \cite{roberts2015localized,hosur2016chaos,hosur2016characterizing,roberts2018operator,qi2018quantum}. If we define the size of an operator as that of its support, a generic operator does not have a definite size but is a linear superposition of operators with different size. Some particular OTOCs\cite{roberts2018operator,qi2018quantum} provides a measure of the average size of an operator, as has been demonstrated in the Sachdev-Ye-Kitaev model\cite{kitaev2015simple,maldacena2016remarks}. More generally, the full distribution of operator size can be studied, which provides more refined information about operator scrambling\cite{hosur2016characterizing,roberts2018operator,qi2018quantum}. Other measures of chaos such as complexity \cite{brown2016complexity,roberts2017chaos} have also been studied.

The OTOC and other measures of chaos are difficult to observe experimentally, since typical experimental observables are time-ordered correlation functions of simple operators.  However, it is possible to measure OTOCs in systems where the flow of time can effectively be reversed by switching the sign of the Hamiltonian ($H\rightarrow -H$) \cite{swingle2016measuring,garttner2017measuring,li2017measuring,wei2018exploring,meier1705j}, as demonstrated in experiments with trapped ions \cite{garttner2017measuring}, solid-state spins \cite{li2017measuring, wei2018exploring}, and Bose-Einstein condensates \cite{meier1705j}.  Whereas the protocols implemented to date allow for probing a restricted set of operators, several proposed approaches to measuring more generic OTOCs---e.g., between spatially separated operators---additionally or alternatively require many-body interferometry  \cite{swingle2016measuring,swingle2018resilience,zhu2016measurement,yao2016interferometric}, which imposes stringent technical demands.  A characterization of scrambling and decoherence by quantum teleportation has been proposed\cite{yoshida2019disentangling} and realized experimentally\cite{landsman2019verified}in a trapped-ion quantum circuit.  Recent theoretical works have shown that measurements of statistical fluctuations and correlations may offer a more convenient route to probing operator scrambling in quantum many-body systems \cite{lewis2019unifying,vermersch2018probing,elben2018statistical}.  Also interpretable as a signature of scrambling are multi-point correlations that have been measured in experiments with ultracold quantum gases \cite{rispoli2018quantum,schweigler2017experimental}.
%\XLQ{(I have added the Yoshida and Monroe refs here.)}

%Very recently, alternative approaches involving randomized measurements have been proposed \cite{vermersch2018probing,elben2018statistical}.

%Several proposals have been made to measure OTOC experimentally\cite{swingle2016measuring,garttner2017measuring,li2017measuring,zhu2016measurement}, which usually involves reversing the Hamiltonian $H\rightarrow -H$.\XLQ{check the statement. distinguish realization papers with proposals} The exception is Ref. \cite{vermersch2018probing,elben2018statistical} which proposed to measure OTOCs and scrambling by statistical correlations between randomized measurements. \footnote{We became aware of Ref. \cite{vermersch2018probing,elben2018statistical} after finishing our work. It is clear that our proposals are based on the same mathematics, although the physical realization seems different.}

In this paper, we propose a method of using straightforward quantum quench experiments to quantify scrambling by directly measuring the growth in operator size. For a system of qudits, such as a quantum spin model, we consider an ensemble of initial states which are direct product states of different qudits. In other words, we consider initial states with zero entanglement between qudits. Then the system is evolved to time $t$ with a Hamiltonian $H$, at which time a physical operator $\hat{O}$ is measured. By preparing the same initial state $\ket{\Psi}$ many times, we can measure the expectation value $O_\Psi(t)\equiv \bra{\Psi}\hat{O}(t)\ket{\Psi}$. By measuring $O_\Psi(t)$ for different initial states $\ket{\Psi}$ in this product state ensemble, we can study its variance $\delta O_\Psi(t)^2$, which characterizes how sensitive the expectation value is to changes in the initial state. Interestingly, this intitial-state sensitivity depends on the size of the operator. For example, in a qubit model if we measure a Pauli operator $\sigma_{x1}$ on a given site $1$, its expectation value is only sensitive to the state of the first qubit, which can vary between $-1$ and $+1$. By comparison, a longer string such as $\sigma_{x1}\sigma_{x2}$ has a smaller variance than $\sigma_{x1}$, since in order for it to take extreme values near $\pm 1$, both $\sigma_{x1}$ and $\sigma_{x2}$ have to be close to $\pm 1$. Based on this intuition, our main result is that the variance of operator expectation value in the product state ensemble is determined by a particular component of the operator size distribution. More specifically, for a traceless operator $\hat{O}(t)$ that is normalized as ${\rm tr}(\hat{O}(t)^2)=D$ ($D$ is the Hilbert space dimension), if it has probability $p_l$ to have size $l$, then we prove the following general result:
\begin{align}
    \delta O(t)^2=\sum_{l=0}^N\frac{p_l}{(d+1)^l}\label{eq:mainresult_intro}
\end{align}
Therefore the bigger operators (with distribution $p_l$ peaking at higher $l$) have smaller variance, although the variance is not determined by the averaged size but the average of the exponentially decaying function $(d+1)^{-l}$. 

Our proposal provides a simple way to measure operator scrambling in systems of cold atoms, trapped ions, or superconducting qubits, where the random initial state ensemble can be prepared. Our result (\ref{eq:mainresult_intro}) is independent of any details of the dynamics of the system, and applies to all operators, although the Heisenberg operators $\hat{O}(t)$ for local observables $\hat{O}(0)$ are those easiest to realize experimentally. We will discuss the essential requirements in the experimental realization of this proposal, and how to take into account errors in state preparation. We will also discuss other information about operator size growth that can be extracted from this setup. We prove that the data $O_\Psi(t)$ as a function of direct product states contains {\it complete information} about operator size distribution. More precisely, for each region $R$, one can denote $p_R$ as the probability that the Heisenberg operator $\hat{O}(t)$ is supported on $R$. The probability $p_R$ for all regions are determined by $O_\Psi(t)$ for the direct product initial state ensemble. This result is given in Eq. (\ref{eq:generalpR}). 

As a variation of our proposal, we also discussed how the OTOC (more precisely the squared commutator) can be measured by preparing an ensemble of fully random initial states (obtained by a Haar random unitary acting on a reference state), rather than random product initial states. For two Hermitian operators $V,W$, we measure the response of $W(t_2)$ generated by a perturbation of $V(t_1)$, with the initial state of the system $\ket{\Psi}$ at $t=0$. This response function is denoted as $C_\Psi(t_2,t_1)$. We show that the second moment of this quantity in the ensemble of random states gives the retarded OTOC:
\begin{align}
    \delta C\left(t_2,t_1\right)^2=-\frac1{D+1}\left\langle \left[W(t_2),V(t_1)\right]^2\right\rangle_{\beta=0}\theta(t_2-t_1)
\end{align}

The remainder of the article is organized as follows. In Sec. \ref{sec:sizedistribution} we define the operator size distribution and discuss some examples. We also define a generating function that is useful for the computation of operator size distribution. In Sec. \ref{sec:proposal} we derive our main result (Eq. (\ref{eq:mainresult_intro}) and discuss some examples. In Sec. \ref{sec:experimental} we discuss the experimental realization of this proposal, focusing on an illustrative example for quantum spin models that can be simulated with cold atoms. In Sec.\ref{sec:discussion} we discuss how more general quantitites can be probed by the same setup, and presented the related proposal that probes the OTOC from the variance of response function. Finally, the conclusion and some open questions are discussed in Sec. \ref{sec:conclusion}.

\section{Operator Size Distribution}\label{sec:sizedistribution}

\subsection{Motivation and Definition}

In classical systems, the phase space coordinate $\xi=\left(p_i,q_i\right)$ satisfies Hamilton equation of motion. Solving the equation of motion with the initial condition $\xi(0)=\xi_0$ determines the trajectory at later time $\xi(t,\xi_0)$. Chaos refers to the fact that $\xi(t,\xi_0)$ depends on $\xi_0$ very sensitively. More quantitatively, the initial value dependence can be quantified by Lyapunov exponents, which are determined by the exponential growth or decay of the singular values of the Jacobian matrix $\frac{d\xi(t,\xi_0)}{d\xi_0}$. In quantum systems, the analog of the coordinate $\xi$ is a quantum operator. For a quantum operator $\hat{O}$, the time evolution is given by the Heisenberg equation $\dot{\hat{O}}(t)=i\left[H,\hat{O}(t)\right]$, which determines $\hat{O}(t)=e^{iHt}\hat{O}(0)e^{-iHt}$. Therefore quantum chaos is characterized by the fact that the Heisenberg picture operator $\hat{O}(t)$ is a ``complicated function" of the $t=0$ operators $\hat{O}(0)$.

However, it is nontrivial to specify what we mean by a complicated function here. Due to unitarity of the quantum mechanics, if we naively pick an orthonormal basis of Hermitian operators $O_\alpha$ (with the orthonormal condition ${\rm tr}\left(O_\alpha O_\beta\right)=\delta_{\alpha\beta}$), we obtain  $\hat{O}_\alpha(t)=U_{\alpha\beta}(t)\hat{O}_\beta(0)$ with $U_{\alpha\beta}(t)$ a unitary matrix, so that there is no analog of Lyapunov exponents. The correct setup instead is to consider a smaller set of orthonormal operators $\hat{S}_n,n=1,2,...,M$, which generate the whole algebra of Hermitian operators. In other words, any Hermitian operator, such as $\hat{O}(t)$, can be expressed as a polynomial of $\hat{S}_n$. We refer to the set $\left\{\hat{S}_n\right\}$ as simple operators. For example, for a one-dimensional single-particle quantum mechanics we can define $\hat{p},\hat{x}$ as simple operators. For a system with $N$ qubits, with the Hilbert space dimension $2^N$, we can define the Pauli operators on each site $\sigma_{an},~a=x,y,z,~n=1,2,...,N$ as simple operators, so that the number of simple operator is $M=3N$. Once the set of simple operator is defined, it is meaningful to distinguish simple versus complicated operators with respect to this set. For example, in the qubit systems, a general traceless Hermitian operator $\hat{O}(t)$ can be expanded in the Pauli basis:
\begin{align}
    \hat{O}(t)=\sum_{n=1}^N\sum_{a=x,y,z}{\psi^{(1)}}_{n}^a\sigma_{an}+\sum_{n_1,n_2,a_1,a_2}{\psi^{(2)}}_{n_1n_2}^{a_1 a_2}\sigma_{a_1n_1}\sigma_{a_2n_2}+...
\end{align}
The components of length $l$ Pauli string has the (real) coefficients \begin{align}
    {\psi^{(l)}}_{n_1n_2...n_l}^{a_1a_2...a_l}=d^{-N}{\rm tr}\left(\hat{O}(t)\sigma_{a_1n_1}\sigma_{a_2n_2}...\sigma_{a_ln_l}\right)\label{eq:operatorWF}
\end{align}
which measures the probability that the operator contains this particular Pauli string. Here $d=2$ is the dimension of Hilbert space on every site. If we normalize $\hat{O}(t)$ as ${\rm tr}\left(\hat{O}(t)^2\right)={\rm tr}(\mathbb{I})=d^N$, the ``operator wavefunction ${\psi^{(l)}}_{n_1n_2...n_l}^{a_1a_2...a_l}$ is also normalized, and we can define the size distribution
\begin{align}
     p_R\left[\hat{O}(t)\right]=\sum_{\alpha_1\alpha_2...\alpha_l}\left({\psi^{(l)}}_{n_1n_2...n_l}^{a_1a_2...a_l}\right)^2\label{eq:distributionDef1}
\end{align}
for each region $R=\left\{n_1n_2...n_l\right\}$.

\begin{figure}[t]
\center
\includegraphics[width=4in]{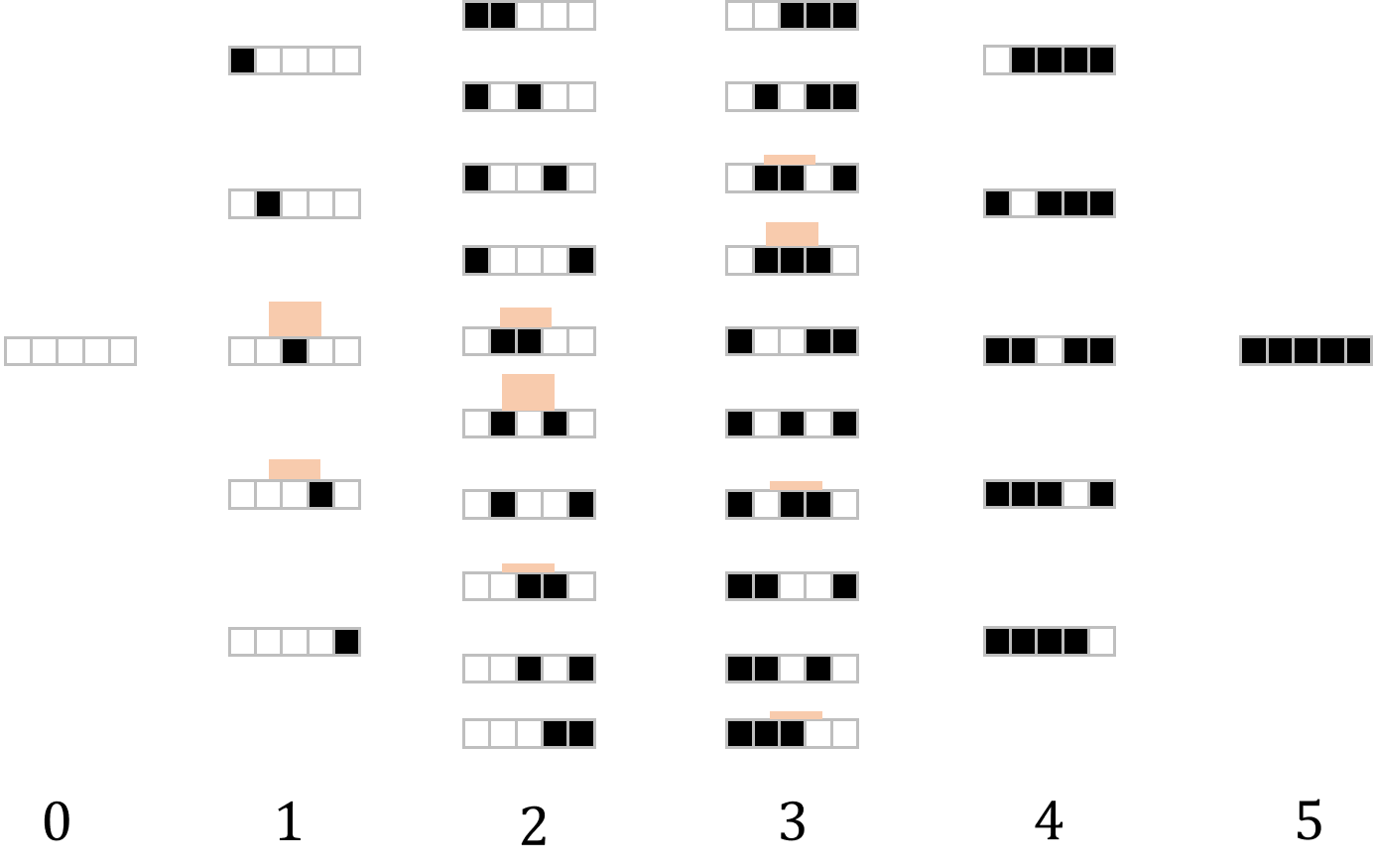}
\caption{Illustration of all possible subregions $R$ on which the Pauli operators $\sigma_{a_1n_1}\sigma_{a_2n_2}...\sigma_{a_ln_l}$ can be supported. The illustration is made for $N=5$ sites, with black boxes representing the sites that an operator is supported on. The orange bars illustrates the absolute value of the corresponding coefficient ${\psi^{(l)}}_{n_1n_2...n_l}^{a_1a_2...a_l}$ for an example operator. The number below each column is the operator size $l$. \label{fig:sizeconfig}}
\end{figure}

One could also sum over all regions with size $l$ (i.e. consisting of $l$ sites) and define the probability the operator has size $l$:
\begin{align}
    p_l\left[\hat{O}(t)\right]=\sum_{|R|=l}P_R\left[\hat{O}(t)\right]\label{eq:distributionDef2}
\end{align}
We use the qubit example for concreteness, but all discussions above trivially generalize to qudit systems with a general $d$-dimensional Hilbert space on each site. For qudits, the Pauli matrices $\sigma_{an}$ are replaced by an orthonormal basis of traceless Hermitian operators in the $d$-dimensional Hilbert space, so that $a=1,2,...,d^2-1$. 

\subsection{Examples}

As a warm up example, we can consider a random Hermitian operator, which means the operator wavefunction ${\psi^{(l)}}_{n_1n_2..n_l}^{a_1a_2...a_l}$ is a random vector in the $d^{2N}-1$ dimensional space of Hermitian operators, obtained by a Haar random rotation from a reference vector. In this case, all Pauli strings of different size appear with equal probability. As a consequence, the probability $p_l$ is simply the probability that a randomly generated Pauli string has length $l$, which is determined combinatorially as
\begin{align}
    p_l=\left(\begin{array}{c}N\\l\end{array}\right)\left(\frac{d^2-1}{d^2}\right)^l\left(\frac1{d^2}\right)^{N-l}\label{eq:distRandom}
\end{align}
The distribution peaks at $l=N\frac{d^2-1}{d^2}$ since the chance each site has a nontrivial operator is $\frac{d^2-1}{d^2}$.

\begin{figure}[t]
\center
\includegraphics[width=5.5in]{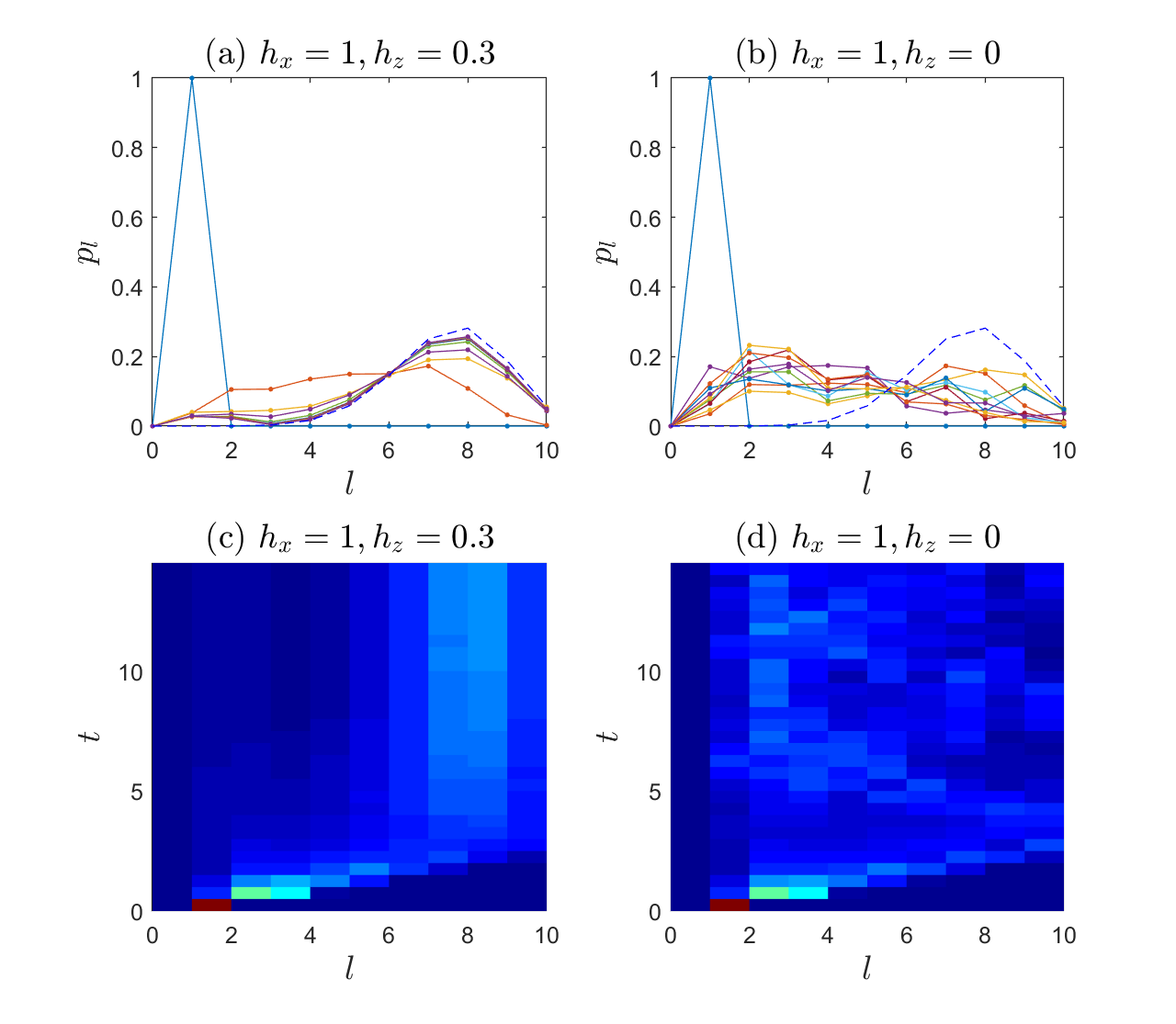}
\caption{The size distribution of $\sigma_{x5}(t)$ for the quantum Ising model of $10$ qubits. The Hamiltonian is chaotic in (a) and integrable (with transverse-field only) in (b). As a comparison, the blue dashed line is the distribution of a random operator given by Eq. (\ref{eq:distRandom}). More details of the time evolution of $p_l\left[\sigma_{x5}(t)\right]$ as a function of $l$ and $t$ is shown in the color maps in (c) and (d). \label{fig:sizedistIsing}}
\end{figure}

For comparison, we can numerically plot $p_l$ for a spin chain model for the time evolution of a Pauli spin operator. Fig. \ref{fig:sizedistIsing} shows the size distribution for $\sigma_{n}^x(t)$ in a quantum Ising chain with the following Hamiltonian:
\begin{align}
    H=\sum_{n=1}^{N-1}J\sigma_{zn}\sigma_{z,n+1}+h_x\sum_{n=1}^N\sigma_{xn}+h_z\sum_{n=1}^N\sigma_{zn}
\end{align}
We see different behavior for integrable chain (with parallel field $h_z=0$) and chaotic chain (with $h_z\neq 0$). For chaotic chain, the size distribution saturates to something similar to that of a random operator, given by Eq. (\ref{eq:distRandom}), which is an evidence of operator scrambling. 

\subsection{An operator identity and the generating function}

\begin{figure}[t]
\center
\includegraphics[width=4.5in]{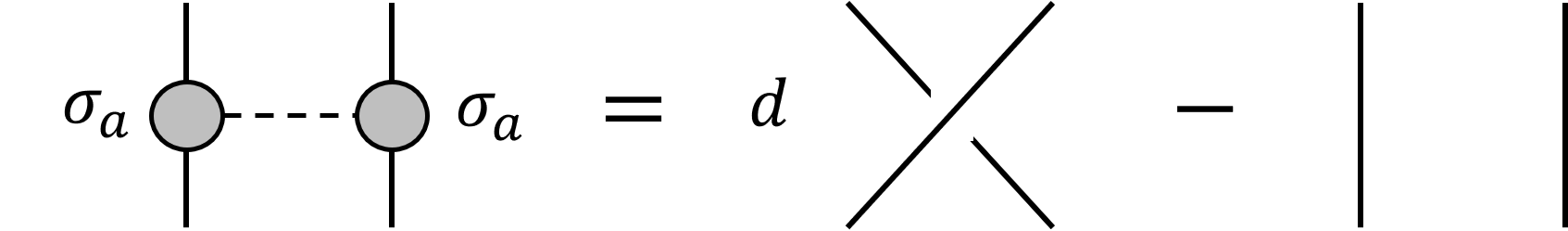}
\caption{Illustration of Eq. (\ref{eq:paulisum}). \label{fig:paulisum}}
\end{figure}

For our discussion later, it is helpful to simplify the expression of operator size distribution (\ref{eq:distributionDef1}) by making use of the orthogonality condition of the operator basis $\sigma_{an}$ in the single-site Hilbert space:
\begin{align}
   {\rm tr}\left(\sigma_{an}\sigma_{bn}\right)&=d\delta_{ab}\nonumber\\
   \sum_{a=1}^{d^2-1}\left[\sigma_{an}\right]_{\alpha\beta}\left[\sigma_{an}\right]_{\gamma\delta}&=d\delta_{\alpha\delta}\delta_{\beta\gamma}-\delta_{\alpha\beta}\delta_{\gamma\delta}
\end{align}
In the second line, we write out the matrix elements of the operator $\sigma_{an}$ as $\left[\sigma_{an}\right]_{\alpha\beta},~\alpha,\beta=1,2,...,d$. The second line is an equation for operators in a doubled Hilbert space $\mathbb{H}\otimes \mathbb{H}$, which can be abbreviated as
\begin{align}
    \sum_{a=1}^{d^2-1}\sigma_{an}\otimes\sigma_{an}=dX_n-\mathbb{I}_n\equiv dW_n\label{eq:paulisum}
\end{align}
with $X_n$ the swap operator that permutes the two copies of the Hilbert space: 
\begin{align}
    \left[X_n\right]_{\alpha\gamma,\beta\delta}=\delta_{\alpha\delta}\delta_{\beta\gamma}\label{eq:swap}
\end{align}
A diagramtic representation of Eq. (\ref{eq:paulisum}) is shown in Fig. \ref{fig:paulisum}.
Using this identity, the sum over $\alpha$'s in Eq. (\ref{eq:distributionDef1}) can be carried explicitly, which leads to
\begin{align}
    p_R\left[\hat{O}(t)\right]&=d^{-2N}{\rm tr}\left[\hat{O}(t)^{\otimes 2}\otimes_{n\in R}\left(\sum_a \sigma_{an}\otimes\sigma_{an}\right)\right]=d^{|R|-2N}{\rm tr}\left[\hat{O}(t)^{\otimes 2}\otimes_{n\in R}W_n\right]\label{eq:pR}
\end{align}

For the computation of $p_l$, it is helpful to introduce a generating function\cite{hosur2016characterizing}:
\begin{align}
    F(z)\left[\hat{O}(t)\right]=\frac{{\rm tr}\left[\hat{O}(t)^{\otimes 2}\otimes_{n=1}^N\left(\frac1d\mathbb{I}_n+zW_n\right)\right]}{{\rm tr}\left(\hat{O}(t)^2\right)}\label{eq:GeneratingFunc}
\end{align}
It is easy to compare with Eq. (\ref{eq:pR}) and conclude
\begin{align}
    F(z)\left[\hat{O}(t)\right]=\sum_{l=1}^N p_lz^l
\end{align}
Using this generating function, $p_l$ can be determined by computing $F(z)$ for $N$ values of $z$ (for example, $z=e^{i\frac{2\pi k}{N+1}},~k=1,2,...,N$), which avoids the sum over $\left(\begin{array}{c}N\\l\end{array}\right)$ different regions. 

For the random operator distribution (\ref{eq:distRandom}), the generating function is
\begin{align}
    F(z)=\left(\frac1{d^2}+\frac{d^2-1}{d^2}z\right)^N
\end{align}

\section{The Quantum Quench Proposal}
\label{sec:proposal}

\subsection{Main result}

\begin{figure}[t]
\center
\includegraphics[width=2.2in]{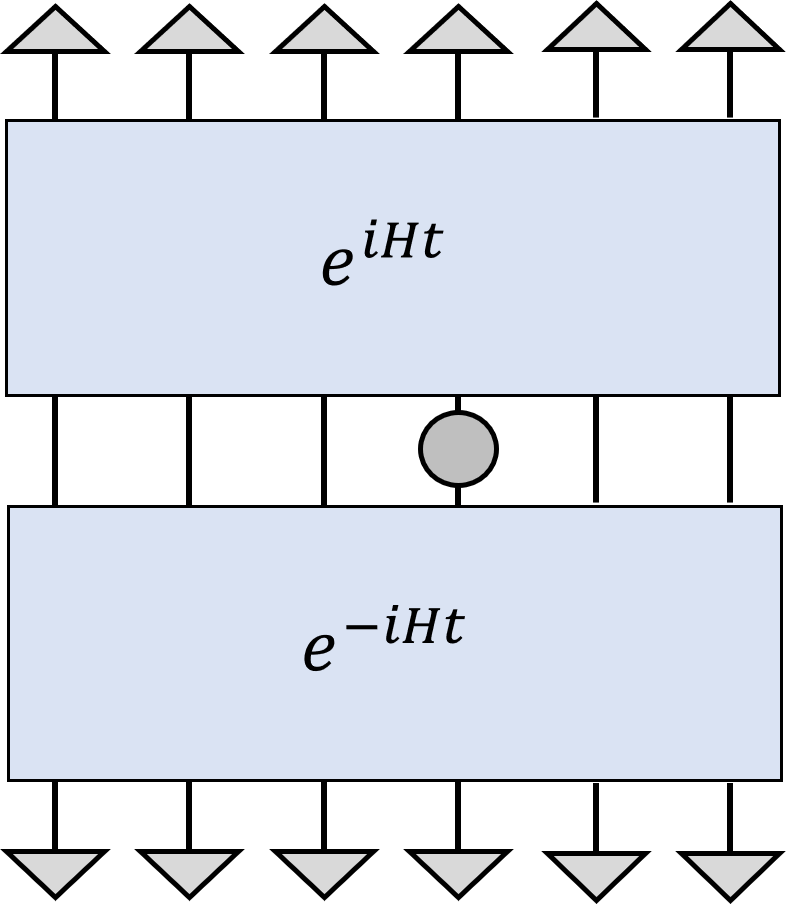}
\caption{Illustration of the quantum quench with unentangled initial states defined in Eq. (\ref{eq:InitialStateEnsemble}).  \label{fig:quench}}
\end{figure}
With this preparation, we now present the main result of this work, which is a proposal that relates the size distribution to physical observables in the setting of quantum quench. 

Consider a system with $N$ qudits, with a Hilbert space dimension of $d^N$. For each qudit, we consider a random state $\ket{\psi_n}$, which is a random vector in the $d$-dimensional Hilbert space with uniform probability distribution. More explicitly, we can choose an arbitrary reference state $\ket{0}$ and define $\ket{\psi_n}=U_n\ket{0}$, with $U_n$ a Haar random unitary operator. For different qudits, $U_n$ are chosen independently, which in the many-body system defines an ensemble of unentangled states
\begin{align}
    \ket{\Psi}\equiv \otimes_{n=1}^N\ket{\psi_n}=\otimes_{n=1}^NU_n\ket{0}\label{eq:InitialStateEnsemble}
\end{align}
Now for a given Hamiltonian $H$, we consider a quantum quench with the initial state $\ket{\Psi}$. For example, in the case of spin $1/2$ qubits, the initial state can be obtained by first preparing a spin-polarized state and then applying a random rotation on each site (see Sec. \ref{sec:experimental}).  Then we evolve the state $\ket{\Psi}$ by a the Hamiltonian $H$. At time $t$ we measure a simple physical operator $\hat{O}$, such as a simple Pauli operator on a single site. By performing this measurement many times (with the same initial state) we can obtain the expectation value of $\hat{O}$:
\begin{align}
    O_\Psi(t)\equiv \bra{\Psi}e^{iHt}\hat{O}e^{-iHt}\ket{\Psi}=\bra{\Psi}\hat{O}(t)\ket{\Psi}
\end{align}
Without losing generality, we always assume $\hat{O}$ is traceless.

In general, the expectation value $O_\Psi(t)$ depends on the initial state. For the initial state ensemble defined in Eq. (\ref{eq:InitialStateEnsemble}), we obtained a probility distribution of $O_\Psi(t)$. Our proposal is that {\it the initial state dependence probes the size distribution of the Heisenberg picture operator $\hat{O}(t)$. } To see this, we consider the second moment of the expectation value:
\begin{align}
    \delta O(t)^2&\equiv\int d\Psi O_\Psi(t)^2-\left(\int d\Psi O_\Psi(t)\right)^2\label{eq:secondmoment}
\end{align}
Here $\int d\Psi=\int \prod_{n=1}^Nd\psi_n$ is the average over all random product states. The integration is carried with a uniform measure that is invariant under $SU(d)$ of each site, and we have chosen the normalization $\int 1 d\Psi=1$. 

\begin{figure}[t]
\center
\includegraphics[width=4.5in]{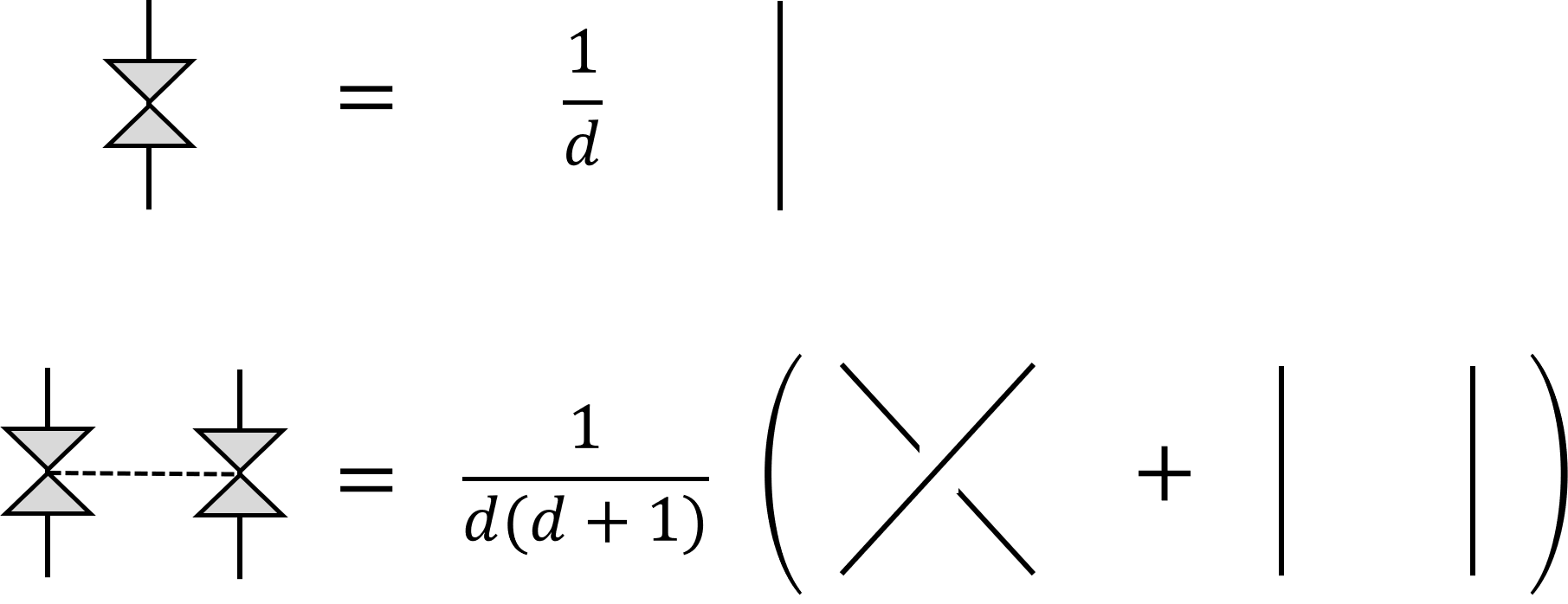}
\caption{Illustration of Eq. (\ref{eq:average}). \label{fig:average}}
\end{figure}

The second moment (\ref{eq:secondmoment}) can be computed using the following two identities:
\begin{align}
    \int d\psi_n\ket{\psi_n}\bra{\psi_n}&=\frac{\mathbb{I}_n}{d},\nonumber\\
    \int d\psi_n\ket{\psi_n}\bra{\psi_n}\otimes \ket{\psi_n}\bra{\psi_n}&=\frac1{d(d+1)}\left(X_n+\mathbb{I}_n\otimes\mathbb{I}_n\right)\label{eq:average}
\end{align}
where $X_n$ is the swap operator defined in Eq. (\ref{eq:swap}). This equation is illustrated diagrammatically in Fig. \ref{fig:average}.  
Using these identities, we obtain
\begin{align}
    \delta O(t)^2&=\frac1{d^N(d+1)^N}{\rm tr}\left[\hat{O}(t)^{\otimes 2}\otimes_{n=1}^N\left(X_n+I_n\right)\right]\nonumber\\
    &\equiv \frac1{d^N(d+1)^N}{\rm tr}\left[\hat{O}(t)^{\otimes 2}\otimes_{n=1}^N\left(W_n+\frac{d+1}dI_n\right)\right]
\end{align}
Compared with the generating function in Eq. (\ref{eq:GeneratingFunc}), we see 
\begin{align}
    \delta {O}(t)^2=\frac{{\rm tr}\left(\hat{O}(t)^2\right)}{d^N}F\left(\frac1{d+1}\right)\left[\hat{O}(t)\right]
\end{align}
In other words, the second moment is determined by the generating function at value $z=\frac1{d+1}$. For simplicity we can take ${\rm tr}\left(\hat{O}(t)^2\right)=d^N$, which is true for Pauli operators. Then 
\begin{align}
    \delta O(t)^2=\sum_{n=1}^Np_n(d+1)^{-n}\equiv \left\langle \frac1{(d+1)^n}\right\rangle\label{eq:mainresult}
\end{align}
which probes a particular linear combination of the operator size distribution $p_n$ for operator $\hat{O}(t)$. The weight of bigger size $n$ decays exponentially, so that roughly speaking, bigger operators have weaker initial state dependence. Intuitively, this is because a complicated operator is highly nonlocal and is less sensitive to a local spin rotation. 

For the random operator with $p_n$ given by Eq. (\ref{eq:distRandom}), one finds 
\begin{align}
    \delta O_{\rm rand}^2=d^{-N},\label{eq:deltaOrand}
\end{align}
which is the inverse of Hilbert space dimension. 

As an example, we show the second moment (computed from $100$ random unentangled initial states) and the direct computation of the right-hand-side of Eq. (\ref{eq:mainresult}) from $p_n$, which shows a good agreement even for the small number of samples. For chaotic systems, the initial state dependence decays with time and saturates around thermalization time, corresponding to the growth and saturation of operator size. For integrable systems, there is more fluctuation in time instead of thermalization. 

\begin{figure}[t]
\center
\includegraphics[width=5.5in]{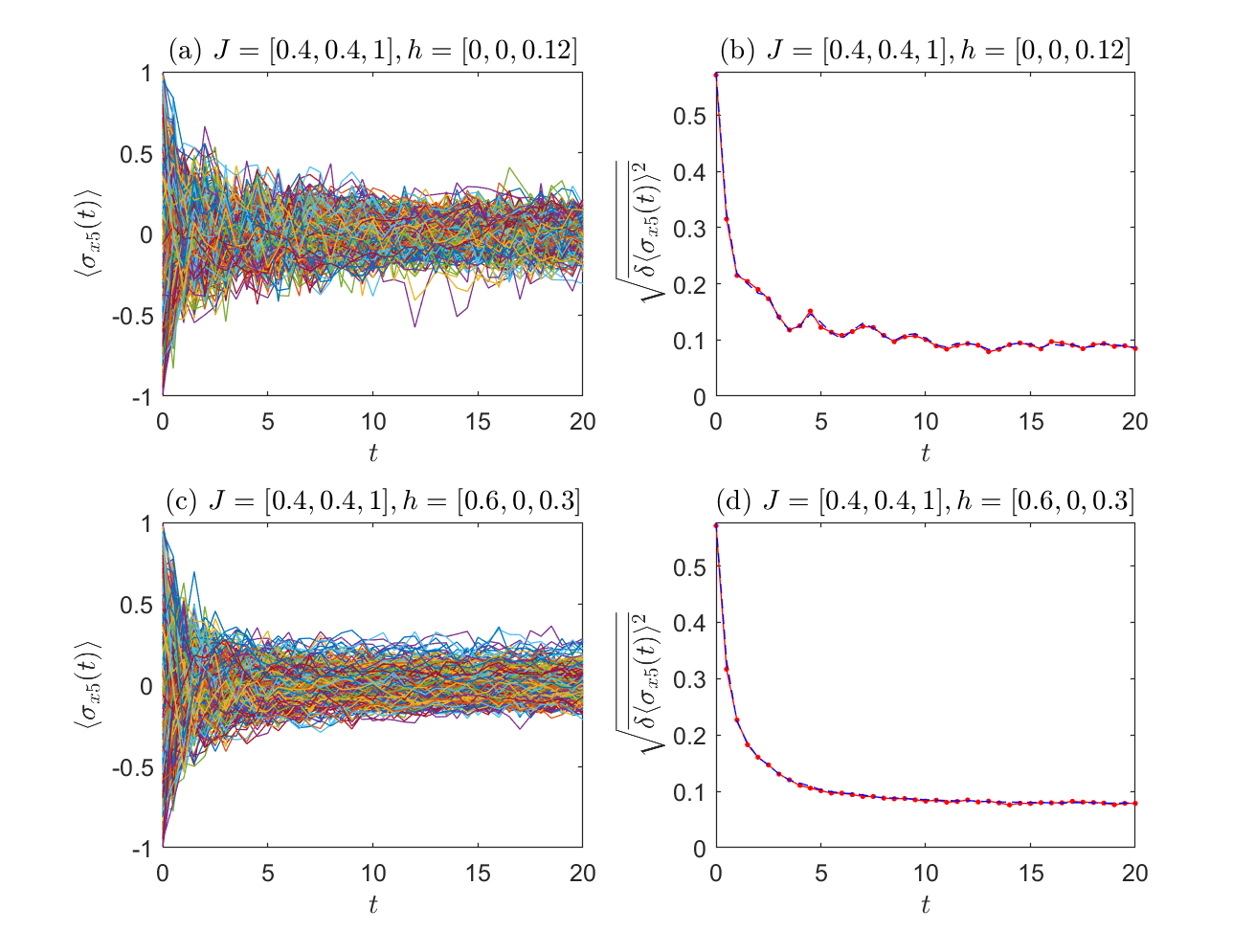}
\caption{(a) (c) The expectation value of $\sigma_{x5}$, the Pauli $x$ operator on the $5$-th site, for a $10$ qubit XXZ model, for $100$ different initial states. The right panels (b) (d) shows the second moment from the $100$ samples (red curve) compared with the direct calculation using Eq. (\ref{eq:mainresult}) (blue dashed lines). The two systems studied are (a) (b) a $U(1)$ symmetric model with only $z$ direction magnetic field; (c) (d) a chaotic model with a generic magnetic field without $U(1)$ symmetry.   \label{fig:XXZ}}
\end{figure}

It should be emphasized that the relation (\ref{eq:mainresult}) applies to arbitrary operators, although experimentally, the operators that are easy to access are Heisenberg evolution of simple operators.

\subsection{Examples}
\begin{figure}[t]
\center
\includegraphics[width=5.5in]{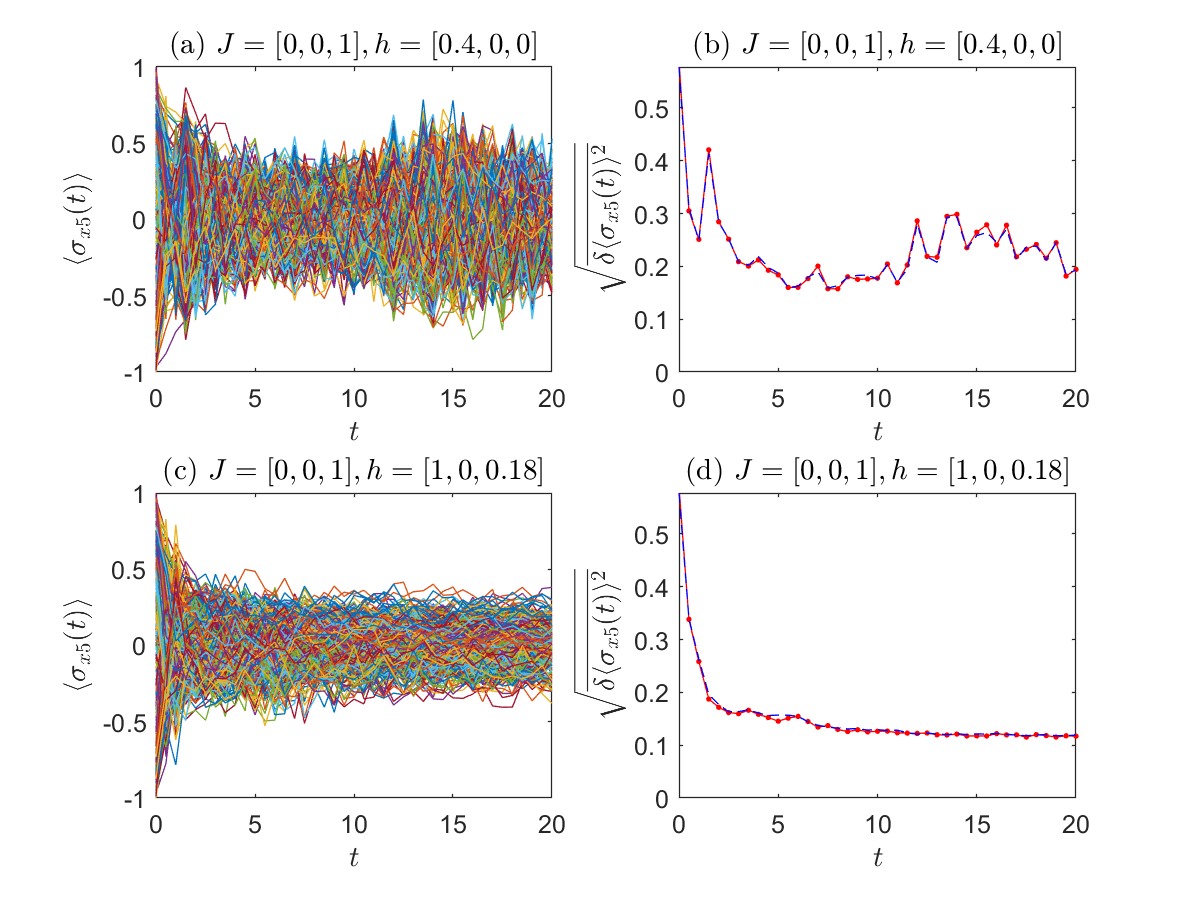}
\caption{The same plot as in Fig. \ref{fig:XXZ} for transverse field ((a) (b)) and chaotic ((c) (d)) Ising model.   \label{fig:Ising2}}
\end{figure}

As an example, we consider a general spin model of the following form:
\begin{align}
    H=\sum_{n=1}^{N-1}\sum_{a=x,y,z}J_a\sigma_{an}\sigma_{a,n+1}+
    \sum_{n=1}^{N}\sum_{a=x,y,z}h_a\sigma_{an}\label{eq:HamXYZ}
\end{align}
In general the model is chaotic, but at particular values it is integrable. For example, if $J_x=J_y,~h_x=h_y=0$, the model is an XXZ model with spin $S_z$ rotation symmetry. In particular, the XY model with $J_z=0, h_x=h_y=0$ can be mapped exactly to a free fermion model. In Fig. \ref{fig:XXZ} we show the comparison between the direct computation of initial state dependent expectation value and the second moment computed from size distribution in Eq. (\ref{eq:mainresult}), for the XXZ model. As expected, the chaotic model has a fluctuation that decreases and saturates at the time of thermalization, while the integrable model (with $h_x=h_y=0$) has more fluctuation. Fig. \ref{fig:Ising2} shows the same computation for the Ising model. Although both transverse field Ising model and XXZ model are solvable, their operator size distribution are still quite different. 

%\noindent{\bf Cavity model.} 
%not sure if we should include it here. The interpretation is not clear. 

\section{Experimental realization}
\label{sec:experimental}

\begin{figure}[t]
\center
\includegraphics[width=4.8in]{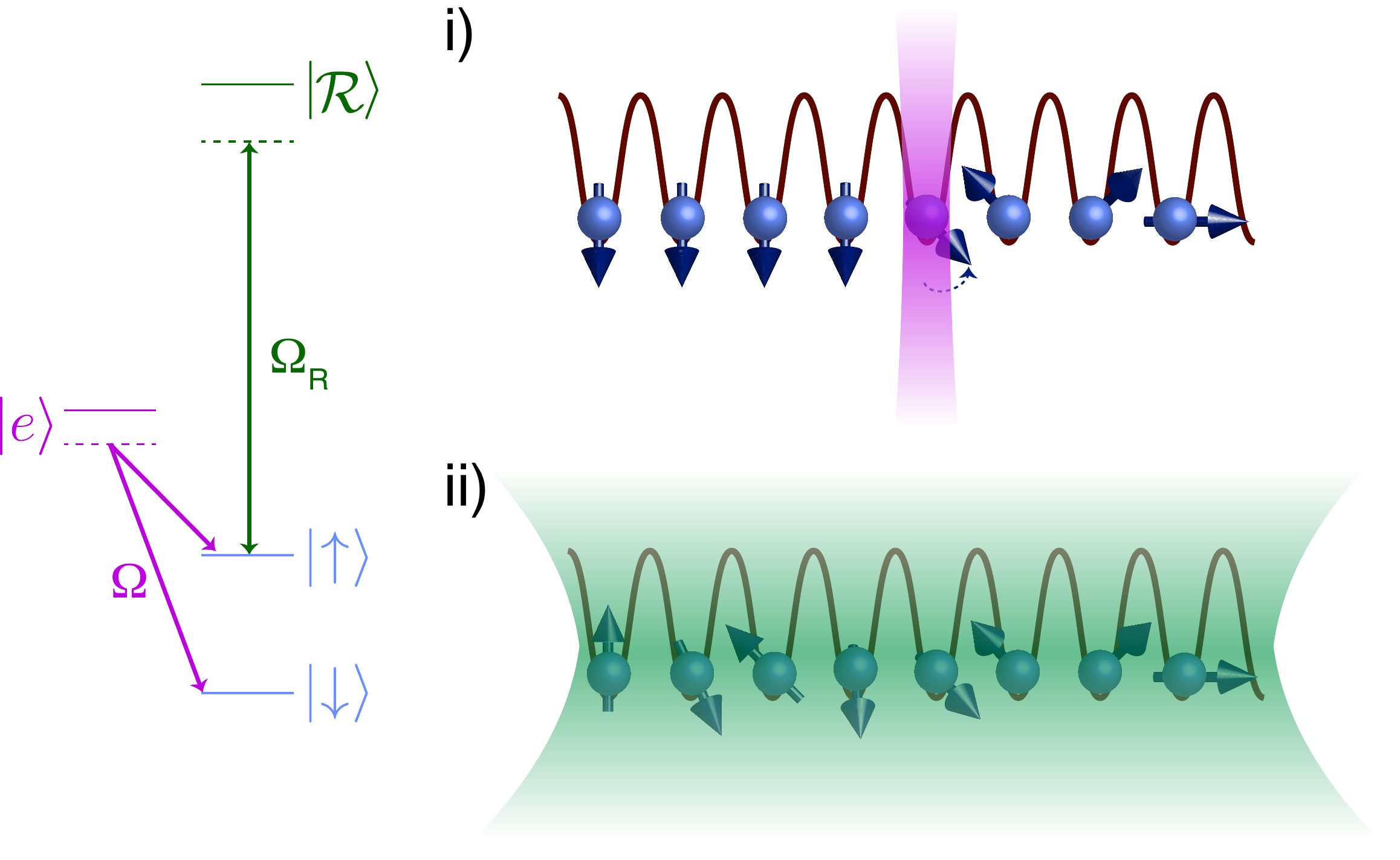}
\caption{A one-dimensional chain of atoms is trapped in an optical lattice and initialized in the spin-down state. i) A focused Raman beam scans from right to left, applying a site-dependent local Hamiltonian $H_n$ to each qubit. ii) The system is uniformly illuminated with light that couples the atoms to Rydberg states to generate spin-spin interactions between the atoms \cite{glaetzle2015designing,van2015quantum,potirniche2017floquet,zeiher2017coherent}. \label{fig:expSetup}}
\end{figure}

Our protocol can be implemented in a variety of physical systems in which spins are encoded in cold atoms, trapped ions, or superconducting qubits.  As an illustrative example, consider a chain of neutral atoms trapped in a microtrap array or optical lattice, where each atom encodes a qubit in a pair of hyperfine ground states (Fig. \ref{fig:expSetup}).  The spin models considered in Sec. \ref{sec:sizedistribution}-\ref{sec:proposal}, including both Ising and Heisenberg spin chains \cite{glaetzle2015designing,van2015quantum,potirniche2017floquet}, can be engineered by coupling the atoms to Rydberg states \cite{labuhn2015realizing,jau2016entangling,zeiher2017coherent,bernien2017probing}. A transverse field is introduced by applying a resonant microwave or Raman coupling between the two hyperfine states, while an effective longitudinal field can be added by detuning this coupling from resonance.  

%Two forms of interactions naturally realizable in this setting are local interactions generated by coupling atoms to Rydberg states \cite{jau2016entangling,labuhn2015realizing,zeiher2017coherent,bernien2017probing}, or all-to-all interactions mediated by photons in an optical cavity \cite{davis2019photon}. \MS{Include figure. Also mention generalization to an array of ensembles instead of an array of individual atoms?}

Arbitrary local initial states can be prepared by first optically pumping all atoms into the spin-down state $\ket{\downarrow}$, and then performing site-dependent rotations with a focused Raman beam controlled by an acousto-optic deflector. In particular, the Raman beam applies a series of local Hamiltonians of the form
\begin{equation}
H_n = \left(\cos\phi_n \sigma_{xn} + \sin\phi_n \sigma_{yn}\right)/2,
\end{equation}
where the phase $\phi_n$ on site $n$ is controlled by the relative phase of two frequency components of the laser field.  Applying the Hamiltonians $H_n$ for times $\theta_n$ yields a state
\begin{align}
\ket{\psi} &= \prod_n e^{-i\theta_n\left[\cos\phi_n \sigma_{xn} + \sin\phi_n \sigma_{yn}\right]/2}\ket{\downarrow}\\
& \propto \prod_n e^{-i\phi_n \sigma_{zn}/2} e^{-i\theta_n \sigma_{xn}/2 }\ket{\downarrow}.
\end{align}
Thus, any desired distribution of initial orientations $\left(\theta_n, \phi_n\right)$ on the Bloch sphere can be generated by a suitable choice of pulse durations $\theta_n$ and phases $\phi_n$.  The above approach furthermore generalizes beyond qubits to qudits: in the case where a larger spin $F>1/2$ is encoded in magnetic sublevels of each individual atom, arbitrary unitary transformations enabling the preparation of random states have been demonstrated in Ref. \cite{anderson2015accurate}.

%The above approach generalizes beyond qubits to qudits, in the case where a larger spin $F>1/2$ is encoded either in magnetic sublevels of each individual atom or in the collective spin of a small ensemble located on each site.  Provided that the Hamiltonian preserves the total spin on each site, random states can easily be prepared by random rotations of initially spin-polarized states.  %Note that random unitaries in more complex multi-level atoms have furthermore been demonstrated in Ref. \cite{}.
%\XLQ{(To realize random states in qudits, it is probably not sufficient to have $SU(2)$ rotations with a random $\theta,\phi$. Actually, $\int_{SU(2)}dUU\ket{0}\bra{0}U^\dagger \otimes U\ket{0}\bra{0}U^\dagger =\sum_{i=1}^MP_i$ with a bigger number of $M$ orthogonal operators $P_i$ (in term of trace norm). For spin $S$ representation, $M=2S+1$. Therefore for $S>1/2$, there are other terms in addition to the $X$ and $\mathbb{I}$ in Eq. (\ref{eq:average}).)}  \MS{If SU(2) rotations are not sufficient, we can either just leave this part out or cite Ref. \cite{anderson2015accurate} on generating arbitrary qudit states within atoms of larger spin.}

We would like to note that our proposal does not require the initial state of each site $\ket{\psi_n}$ to have a totally uniform (i.e. $SU(d)$ symmetric) probability distribution. Since our proposal has only used the second moment of operator expectation values, we only require the initial state ensemble to satisfy Eq. (\ref{eq:secondmoment}). In other words, the initial state ensemble is only required to form a spherical $2$-design\cite{delsarte1991spherical}. It is sufficient to prepare an ensemble of finite number of initial states $\ket{\psi_n^{(i)}}$ with probability $p_i$, for $i=1,2,...,M$, as long as the following conditions are satisfied:
\begin{align}
    \sum_ip_i\ket{\psi_n^{(i)}}\bra{\psi_n^{(i)}}&=\frac1d\mathbb{I}_n\nonumber\\
    \sum_ip_i\ket{\psi_n^{(i)}}\bra{\psi_n^{(i)}}\otimes\ket{\psi_n^{(i)}}\bra{\psi_n^{(i)}}&=\frac1{d(d+1)}\left(X_n+\mathbb{I}_n\otimes \mathbb{I}_n\right)\label{eq:2design}
\end{align}

One main challenge for the experimental realization of our proposal is the accurate measurement of expectation value $\langle \hat{O}(t)\rangle$, which requires preparing and measuring the same initial state multiple times. According to the central limit theorem, the measured value of $O(t)$ in $k$ measurements approach the expectation value with accuracy $\Delta \sim k^{-1/2}$. To verify our proposal of initial state dependence, we need 
\begin{align}
    \Delta^2\ll \delta\left[\hat{O}(t)\right]^2
\end{align}
which requires
\begin{align}
    k\gg \frac1{\delta\left[\hat{O}(t)\right]^2}
\end{align}
(Here we have assumed the operator norm of $\hat{O}$ to be order $1$.) If $\hat{O}(t)$ approaches a random operator in long time, according to Eq. (\ref{eq:deltaOrand}) one has to require $k\gg d^N$. The exponential growth makes the experiment difficult at late time for a large system. This is similar to the measurement of second Renyi entropy\cite{daley2012measuring,islam2015measuring,elben2018renyi,brydges2019probing}. Nevertheless, for a given number of experiments $k$, our proposal can still give information about operator size distribution at early time.

\section{Further discussion}\label{sec:discussion}

\subsection{Effect of state preparation error}

Since it is difficult to prepare the same state many times experimentally, it is natural to study what happens if the state preparation is not perfect. Consider the preparation of an initial state $\ket{\psi_n}$ for $k$ times. As a simple model of the preparation error, let us assume the states prepared are actually $\sqrt{1-\epsilon}\ket{\psi_n}+\sqrt{\epsilon}\ket{\phi_i},~i=1,2,...,k$. If we assume $\ket{\phi_i}$ are independent (normalized) random states, with a uniform distribution, in the large $k$ limit the first and second moment in Eq. \ref{eq:secondmoment} is replaced by
\begin{align}
    \overline{\ket{\psi_n}\bra{\psi_n}}&=\frac{\mathbb{I}_n}{d}\nonumber\\
    \overline{\ket{\psi_n}\bra{\psi_n}\otimes \ket{\psi_n}\bra{\psi_n}}&=\frac{(1-\epsilon)^2}{d(d+1)}\left(X_n+\mathbb{I}_n\otimes\mathbb{I}_n\right)+\left(2\epsilon-\epsilon^2\right)\frac{\mathbb{I}_n\otimes\mathbb{I}_n}{d^2}\nonumber\\
    &=\frac{(1-\epsilon)^2}{d(d+1)}W_n+\frac{\mathbb{I}_n\otimes\mathbb{I}_n}{d^2}
\end{align}
Use this identity one can easily see that the width of distribution is modified to
\begin{align}
    \delta O(t)^2=\sum_{n=1}^Np_n\left(\frac{(1-\epsilon)^2}{d+1}\right)^n\equiv F\left(\frac{(1-\epsilon)^2}{d+1}\right)\left[\hat{O}(t)\right]
\end{align}
Therefore the relation between initial state dependence and size distribution remains qualitatively the same, but the fluctuation decreases with increasing error $\epsilon$. If the distribution of the error is not uniform, the fluctuation will involve non-universal contributions which cannot be related to operator size distribution. 

\subsection{Probing more general quantities}\label{sec:OTOC}

Once the expectation value $O_\Psi(t)=\langle \Psi|\hat{O}(t)|\Psi\rangle$ is measured for the product state ensemble, many other features of the operator size growth can be obtained in addition to the second moment. For example, for each region $R$, we can average $O_\Psi(t)$ over $\ket{\psi_n}$ for sites $n\notin R$. This leads to
\begin{align}
    O_\Psi^R(t)\equiv \int \prod_{n\notin R}d\psi_n\bra{\Psi}\hat{O}(t)\ket{\Psi}={\rm tr}\left[\hat{O}(t)\prod_{n\in R}\ket{\psi_n}\bra{\psi_n}\otimes\otimes_{n\notin R}\frac{\mathbb{I}_n}{d}\right]
\end{align}
The initial state dependence of $O_\Psi^R(t)$ can be studied in the same way as $\delta O(t)^2$ we studied earlier, which is related to the size distribution of $\hat{O}(t)$. Due to the partial trace on $\overline{R}$, now the only terms in $\hat{O}(t)$ that contribute to initial state dependence are those supported on $R$ or its subset. The second moment is given by
\begin{align}
    \delta {O^R(t)}^2=\sum_{S\subseteq R}p_S(d+1)^{-|S|}
\end{align}
This quantity provides information about the spatial distribution and locality of $\hat{O}(t)$. At later time of a chaotic time evolution, most terms in $\hat{O}(t)$ have a large support, so that we expect $\delta O_R^2$ to be small for regions smaller than half the system size. For a random operator, one can see
\begin{align}
    \delta {O^R_{\rm rand}}^2=\frac1{d^{|S|}}\frac1{d^{2(N-|S|)}}=\frac{d^{|S|}}{d^{2N}}
\end{align}

In principle, from the region-dependent variation $\delta {O^R(t)}^2$ of different regions, one can extract the full distribution $p_R$ for all regions:
\begin{align}
    p_R&=\frac{D_R}{D^2}{\rm tr}\left(\hat{O}(t)^{\otimes 2}W_R\right)\nonumber\\
    &=\frac{D_R}{D^2}{\rm tr}\left[\hat{O}(t)^{\otimes 2}\otimes_{x\in R}\left(d(d+1)\overline{\ket{\psi_x}\bra{\psi_x}^{\otimes 2}}-\frac{d+1}d\mathbb{I}_x\right)\right]\nonumber\\
    &=\frac{D_R^2(d+1)^{|R|}}{D^2}{\rm tr}\left[\hat{O}(t)^{\otimes 2}\otimes_{x\in R}\left(\overline{\ket{\psi_x}\bra{\psi_x}^{\otimes 2}}-\overline{\ket{\psi_x}\bra{\psi_x}}^{\otimes 2}\right)\right]\nonumber\\
    &=(d+1)^{|R|}\sum_{Q\subseteq R}\left(-1\right)^{|R|-|Q|}{\delta {O^Q}(t)^2}\label{eq:generalpR}
\end{align}
In practice, computing $p_R$ is a difficult task for a big region $R$ because of the sum over $2^{|R|}$ subsets of $R$ with alternating signs.

\subsection{Measuring the OTOC as a byproduct}

In this section, we discuss a different but related setup which can be used to measure the OTOC. If instead of the random product state ensemble, we consider random initial states, with a uniform probability measure in the $d^N$ dimensional Hilbert space, then we have
\begin{align}
    \int d\Psi\ket{\Psi}\bra{\Psi}&=\frac{\mathbb{I}}{D}\nonumber\\
    \int d\Psi \ket{\Psi}\bra{\Psi}\otimes \ket{\Psi}\bra{\Psi}&=\frac{X+\mathbb{I}\otimes\mathbb{I}}{D(D+1)}
\end{align}
with $D=d^N$ the Hilbert space dimension. Here $X$ is the swap operator acting on all sites. If we study the initial state dependence of an operator $\hat{O}(t)$, we obtain the variance
\begin{align}
    \delta O(t)^2&=\int d\Psi\bra{\Psi}\hat{O}(t)\ket{\Psi}^2-\left(\int d\Psi\bra{\Psi}\hat{O}(t)\ket{\Psi}\right)^2\nonumber\\
    &=\frac{{\rm tr}\left(\hat{O}(t)^2\right)}{D(D+1)}\equiv \frac1{D+1}\langle \hat{O}^2\rangle_{\beta=0}
\end{align}
Compared with the case of initial product state ensemble, the initial state dependence in this case does not probe the operator size of $O(t)$. However, this proposal can be generalized by measuring a response property for each initial state $\ket{\Psi}$. If we apply a perturbation $\epsilon V$ at time $t_1$, and measure the response of operator $W$ at a later time $t_2$, the response coefficient is given by the retarded Green's function
\begin{align}
    C_\Psi(t_2,t_1)=-i\bra{\Psi}\left[W\left(t_2\right),V\left(t_1\right)\right]\ket{\Psi}\theta(t_2-t_1)
\end{align}
with $\theta(t)$ the step function that equals $1$ at $t>0$ and vanishes otherwise. 

Therefore it is easy to see that the variance of $C_\Psi(t_2,t_1)$ has the form
\begin{align}
    \delta C(t_2,t_1)^2=-\frac1{D+1}\langle \left[W(t_2),V(t_1)\right]^2\rangle_{\beta=0}\theta(t_2-t_1)\label{eq:deltaC2}
\end{align}
which is determined by the infinite temperature OTOC between this pair of operators. In other words, the squared commutator measures the initial-state dependence of a linear response function for totally random initial states. 

Although this proposal also probes the operator size distribution of operators $W$ and $V$ (through commutator square), it is more difficult to realize experimentally than our main proposal using product initial states, due to two reasons. Firstly, preparing the random initial state ensemble is much more difficult than preparing the random product state ensemble. Secondly, the quantity (\ref{eq:deltaC2}) is suppressed by the exponentially small factor $\frac1{D+1}$ for all time, while $\delta \langle O(t)\rangle ^2$ for the random product states is much bigger at early times. 

\section{Conclusion}
\label{sec:conclusion}

In conclusion, we have proposed a general measure of operator size distribution based on the initial-state dependence of one-point functions. Compared to previous proposals, the main advantages of our proposal are its generality and relative ease of implementation. Our method can be implemented in any quantum quench experiment as long as the direct product state ensemble can be prepared, and simple operator expectation values at a later time can be measured. The easiest operators to measure are Heisenberg operators $\hat{O}(t)$ for local operators $\hat{O}$, but all our results relating the size distribution and initial-state dependence apply to generic operators.  In spirit, our results relate quantum chaos to initial state dependence of the dynamics, in a similar way as the relation of classical chaos and initial condition sensitivity.

While the examples in this paper have focused on applications to nearest-neighbor spin chains, our method can be implemented in a variety of different experimental contexts allowing for the study of a wide range of Hamiltonians.  Examples include long-range power-law interactions realizable with trapped ions \cite{jurcevic2014quasiparticle,richerme2014non} or Rydberg atoms \cite{labuhn2015realizing,jau2016entangling,zeiher2017coherent,bernien2017probing}, or non-local photon-mediated interactions in optical cavities \cite{davis2019photon} enabling the realization of Sachdev-Ye-like models \cite{strack2011dicke,gopalakrishnan2011frustration}.  While spin systems offer a particularly convenient means of preparing random initial states via random rotations, our protocol could also be extended to Hubbard models of ultracold atoms in optical lattices by adopting methods proposed in Ref. \cite{elben2018renyi} for the implementation of random unitaries.  Notably, such systems would allow for probing how operator growth is affected by many-body localization \cite{schreiber2015observation,choi2016exploring,lukin2019probing,rispoli2018quantum}.

If different initial states can be prepared in the experimental setup, our method also offers substantial flexibility in defining the measure of operator size distribution. For example, we can consider an ensemble of initial states 
\begin{align}
    \ket{\Psi}=\ket{\psi_{12}}\otimes\ket{\psi_{34}}\otimes...\otimes\ket{\psi_{N-1,1}}\label{eq:cluster}
\end{align}
with $\ket{\psi_{n,n+1}}$ random states of dimension $d^2$ for the two sites $n,n+1$. The variance of $O_\Psi(t)=\bra{\Psi}\hat{O}(t)\ket{\Psi}$ in this ensemble will probe the size distribution $\tilde{p}_l$ of operator $\hat{O}(t)$ with a different definition of size. For example denote $\sigma_{\alpha 1}$ and $\sigma_{\alpha 2}$ as Pauli operators on sites $1,2$ respectively. Then in the original case of product state ensemble $\sigma_{\alpha 1}\sigma_{\beta 2}$ is  considered as a size $2$ operator. Now in the modified ensemble (\ref{eq:cluster}), the variance of expectation value is probing a ``coarse-grained size'' where $\sigma_{\alpha 1},~\sigma_{\beta 2},~\sigma_{\alpha 1}\otimes\sigma_{\beta 2}$ are all considered as size $1$ operators. Therefore the initial state ensemble defines the ``size units''. A more entangled initial state ensemble probes a more coarse-grained size measure. 

If the initial state ensemble is not totally random--- for example, for a qubit system, if the spin has higher chance to be along $z$ direction than $x,y$---then the variance of $O_\Psi(t)$ will not be determined by the size distribution alone, but will depend on more complicated multipoint functions. It is an open question how to estimate the effects of an imperfect initial state ensemble, and whether in some cases one can still find approximate relations or inequalities between the initial state dependence and operator size distribution.

Another open question is whether there are interesting information in higher moments of the initial state dependence $O_\Psi(t)$. The relation to size distribution only used the second moment, and the higher moments may provide probes to other features of operator scrambling. 

{\bf \noindent Acknowledgement.} We would like to thank Daniel Bulmash, Yingfei Gu, Pavan Hosur, Greg Bentsen, and Eric Cooper for helpful discussions. This work was supported by the DOE Office of Science, Office of High Energy Physics, the grant DE-SC0019380 (XLQ, MSS, AP), the Simons Foundation (XLQ), the Research Corporation Cottrell Scholar Program (MSS), the National Science Foundation (ED), and the Hertz Foundation (ED).

\bibliography{refs}\bibliographystyle{jhep}

%\end{thebibliography}

\end{document}